\documentclass[preprint]{emulateapj}
\newcommand{\lya}{\ion{H}{1}~Ly$\alpha$}
\newcommand{\hmol}{\hbox{H$_2$}}
\newcommand{\Msun}{\hbox{$M_\odot$}}
\newcommand{\Mdot}{\hbox{$\dot M$}}
\newcommand{\Msunyr}{\hbox{$M_\odot\,$year$^{-1}$}}
\newcommand{\altion}[2]{\textup{#1}\,\textsc{#2}}
\newcommand{\vel}[1]{$-$#1 \kms{}}

\def\hst{{\it HST}}
\def\vlt{{\it VLT}}
\def\vlti{{\it VLTI}}
\def\ec{$\eta$\,Car}
\def\kms{km~s$^{-1}$}
\def\cm{cm$^{-1}$}
\def\lm{$\lambda$}

\begin{document}

\title{Eta Carinae across the 2003.5 Minimum: Analysis in the visible and near infrared spectral region}

\author{K.~E. Nielsen\altaffilmark{2,3}, G. Vieira~Kober\altaffilmark{2,3}, K. Weis\altaffilmark{4}, T. ~R. Gull\altaffilmark{3}, O. Stahl\altaffilmark{5}, D.~J. Bomans\altaffilmark{4}}

\shorttitle{Analysis in the visible and near infrared spectral region} 
\shortauthors{Nielsen et\,al.}

\altaffiltext{2}{Catholic University of America, Washington, DC 20064}
\altaffiltext{3}{Astrophysics Science Division, Code 667, Goddard Space Flight Center, Greenbelt, MD 20771}
\altaffiltext{4}{Astronomisches Institut, Ruhr-Universit\"at Bochum, Universit\"atsstrasse 150, D-44780 Bochum, Germany}
\altaffiltext{5}{ZAH, Landessternwarte Heidelberg-K\"onigstuhl, D-69117 Heidelberg, Germany}

\email{krister.nielsen@nasa.gov}

\begin{abstract} 
We present an analysis of the visible through near infrared spectrum of \ec\ and its ejecta obtained during the "\ec\ Campaign with the Ultraviolet and Visual Echelle Spectrograph (UVES) at the ESO \textit{Very Large Telescope} (\vlt)".  This is a part of larger effort to present a complete \ec\ spectrum, and extends the previously presented analyses with the \textit{Hubble Space Telescope}/Space Telescope Imaging Spectrograph (\hst/STIS) in the  UV (1240$-$3159~\AA) to 10,430~\AA. The spectrum in the mid and near UV is characterized by the ejecta absorption. At longer wavelengths, stellar wind features from the central source and narrow emission lines from the Weigelt condensations dominate the spectrum. However, narrow absorption lines from the circumstellar shells are present. This paper provides a description of the spectrum between 3060 and 10,430~\AA, including line identifications of the ejecta absorption spectrum, the emission spectrum from the Weigelt condensations and the P-Cygni stellar wind features. The high spectral resolving power of \vlt/UVES enables equivalent width measurements of atomic and molecular absorption lines for elements with no transitions at the shorter wavelengths. However, the ground based seeing and contributions of nebular scattered radiation prevent direct comparison of measured equivalent widths in the \vlt/UVES and \hst/STIS spectra. Fortunately, \hst/STIS and \vlt/UVES have a small overlap in wavelength coverage which allows us to compare and adjust for the difference in scattered radiation entering the instruments' apertures. This paper provides a complete online \vlt/UVES spectrum with line identifications and a spectral comparison between \hst/STIS and \vlt/UVES between 3060 and 3160 \AA. 
\end{abstract}


\section{Introduction} 
Eta Carinae (\ec) is a massive Luminous Blue Variable \citep[LBV,][]{Conti84a}, that enriches the interstellar medium through major mass eruptions as it transitions from hydrogen burning to helium core nuclear energy production \citep{Humphreys94a}. During the LBV phase, the star, for a relatively brief time interval,  brightens as a consequence of a major lift-off of the stellar atmosphere's outer layers. A 60 $M_\odot$ star may lose up to 10 $M_\odot$ while on the main sequence, but its pre-supernova mass is thought to be about 20 $M_\odot$ \citep{Smith06b}. The LBV phase is likely an evolutionary stage which most massive stars encounter during their lifetime, before exploding as a supernova \citep{Meynet03a}. Growing evidence supports that some core-collapse supernovae have LBV progenitors \citep[e.g.][]{Smith07c, GalYam07a}, enhancing the cosmological importance of the LBVs. \\
\indent Eta Car has a past with LBV characteristic events. The ejecta, formed during the Great Eruption in the 1840s and the less dramatic event in the 1890s, modifies \ec's spectrum. In the 1840s more than 12 $M_\sun$ of material was ejected \citep{Smith03a} and formed the bipolar Homunculus with an intervening disk \citep{Davidson01a}. In the 1890s an ionized bipolar structure, the Little Homunculus, formed interior to the Homunculus with a total mass of $\sim$0.5 $M_\sun$ \citep{Ishibashi03a}. The Homunculus reflection nebula covered approximately 10\arcsec $\times$ 18\arcsec\ on the sky in 2000 \citep{Morse01, Davidson01a}, while the Little Homunculus is about 4\arcsec\ in extent \citep{Ishibashi03a}. The Weigelt condensations \citep{Weigelt86} are bright emission structures currently located 0\farcs2 to 0\farcs3 from the stellar system. They date back to the event in the 1890s based upon proper motion measurements \citep{Davidson95a, Smith04a}. A peculiar ionized region, the Strontium filament, lies within the disk \citep{Zethson01a} and  is excited by \ion{H}{1} Balmer continuum radiation filtered by singly ionized iron \citep{Hartman04}. The Strontium filament's spectrum shows lines of allowed and forbidden \ion{Sr}{2} lines but is dominated by \ion{Ti}{2}, \ion{Ni}{2} and \ion{Fe}{1}. Neither \ion{H}{1} nor \ion{He}{1}, and only weak \ion{O}{1} and \ion{C}{1} emissions are observed. Outside the Homunculus, a larger nebula of thin filaments, the outer ejecta, is observed. The outer ejecta is expanding with velocities up to 2000 \kms\ \citep{Weis04a} with a total mass of 2$-$4 \Msun (K. Weis, 2008, private communication). \\
\indent Studies of the nebular emission revealed a nitrogen-enhanced but carbon- and oxygen-starved system \citep{Davidson86a, Dufour97a}. \citet{Smith04d} observed depleted oxygen abundances but enhanced nitrogen in the ejecta located immediately outside the Homunculus, while the ejecta at larger distance has solar composition. \citeauthor{Smith04d} concluded that the ejecta further away from the central object was thrown long before the creation of the Homunculus and was less CNO processed. \citet{Verner05a} suggested that oxygen and carbon are depleted by two orders of magnitude in the Weigelt condensations. The massive star that ejected this nitrogen-rich material was in its hydrogen burning evolutionary stage. Massive stars ($>$40\Msun) are shown to have enhancements of nitrogen at the expense of carbon and oxygen. This is  an abundance characteristic more pronounced due to mixing caused by stellar rotation \citep{Meynet00a,Meynet03a}. The remaining oxygen is tied up in corundum and simple molecules, leaving much of the metals in their gaseous atomic phase \citep{Chesneau05a}. The dust in the ejecta surrounding \ec\ appears to be primarily in a dusty torus and in the outer Homunculus \citet[][and references therein]{Smith03a, Smith02}. However, the evidence is sparse as to if carbon is bound in the dust in the system. Furthermore, the properties of dust formation in a nitrogen-rich chemistry is rarely addressed.\\
\indent Eta Car has a spectroscopic period, initially discovered  through variations in \ion{He}{1} $\lambda$10830 integrated flux \citep[][and references within]{Damineli96a}. The variation in excitation is accompanied by eclipse-like minima observed with \textit{UBV} and \textit{BVR} photometry \citep{vanGenderen03}, in near infrared \textit{JHKL} photometry \citep{Whitelock04} and in X-ray brightness \citep{Ishibashi99a, Corcoran05a}.  \citet[][ and references therein]{Corcoran05a} fine-tuned \ec's period to 2024$\pm$2 d (5.54 year) based on variability in the X-ray emission. Recently, \citet{Damineli08a} showed that the visible spectrophotometry and X-ray variability give consistent results. Most, if not all, observations can be explained by the presence of a companion star in a highly eccentric orbit with a 5.54 year period. A discussion about the binarity of the system can be found in several papers including \citet{Nielsen07a}, where the binary solution is compared to a single star model that attributes the spectroscopic variability from a latitudinal dependent stellar wind \citep{Smith03b}. We adopted the binary solution for the spectral analysis presented in this paper, where the periastron passage represents the system's spectroscopic low-state and the time outside the periastron passage is the spectroscopic high-state.   \\
\indent Much attention has been devoted to find direct spectral evidence for the companion star (\ec~B) or signatures from the interacting winds.  The X-radiation was demonstrated to originate from the wind-wind interface region \citep{Ishibashi99a, Corcoran01a} and was modeled as interaction between a primary star (\ec~A) with a dense wind (\Mdot $\sim$2.5$\times$10$^{-4}$ \Msunyr, $v_\infty$=500 \kms)  and a fast companion wind \citep[\Mdot $\sim$10$^{-5}$ \Msunyr, $v_\infty$=3000 \kms; ][] {Pittard02}. \citet{Steiner05, Stahl05} detected \ion{He}{2} in the spectrum of \ec\ indicating the presence of a hot companion. The \ion{He}{2} emission was demonstrated by \citet{Martin05} to originate from the wind-wind interface where highly excited lines normally not observed in the wind spectrum of a 15,000~K star could form. \citet{Nielsen07b}  traced, with \hst/STIS spectra, \ion{He}{1} lines formed in the wind-wind interface over the 5.54 year spectroscopic period to further analyze the ionization structure of the wind and derive \ec~B's orbital parameters. The spectrum of the Weigelt condensations during the broad maximum is consistent with the presence of a 37,000 K object, implying a companion star of O or WR-type \citep{Verner05a}. \citet{Iping05a} found evidence for a hot companion by the far-UV flux behavior, but could not characterize it any further as the far-UV spectrum is modified by \ec~A's extended wind and the circumstellar material. All evidence favor a scenario where \ec\ is a massive binary. \\
\indent  \vlti/VINCI interferometry \citep{vanBoekel03} infered that \ec~A is a prolate spheroid, extending along the polar axis of the Homunculus, consistent with a polar wind with variable velocity during the spectroscopic high-state that decreased during the 1998.0 low-state \citep{Smith03b}. The results from \citeauthor{vanBoekel03} agree well with more recent \vlti/AMBER interferometry, using visibility functions of  \ion{He}{1} at 2.059 $\mu$m and \ion{H}{1} Br$\gamma$ \citep{Weigelt07a}. \\
\indent  Multiple \hst\ programs monitored the spectroscopic variations in the wavelength range 1175$-$10,300~\AA\ from 1998.0 to 2004.3. The \hst/STIS data provides invaluable information about the character and geometry of \ec\ and its ejecta, including the geometry of the Homunculus \citep{Davidson01a} and the Little Homunculus \citep{Ishibashi03a}. The initial spectral analysis using data obtained with \hst/STIS resulted in atlases for the complete medium \citep[R$\sim$30,000, 1240$-$2360 \AA;][]{Nielsen05a} and high \citep[R$\sim$110,000, 2380$-$3159 \AA;][]{Gull06a} resolution STIS MAMA wavelength region in  the sightline towards the star. The high resolution modes were exclusively used for the analysis of the circumstellar absorption shells, since the narrow line absorptions  not are observable in the low resolution (R$\sim$8000) STIS CCD spectrum.  Spectral atlases, using the \hst/STIS CCD spectrum (R$\sim$8000), for the Strontium filament and the Weigelt condensations are found in \citet{Hartman04} and  \citet{Zethson01b}, respectively. The Weigelt D UV spectrum was, with limited wavelength coverage, investigated with high resolution STIS MAMA spectra \citep{Nielsen07b}. These atlases serve as a platform for further studies of \ec's ejecta and circumstellar material around other objects such as gamma ray burst progenitors \citep{Prochaska06a, Vreeswijk07a} or other LBVs (Nielsen et\,al., in prep).

\begin{figure}[ht]
\centering
  \includegraphics[angle=0, scale=0.315]{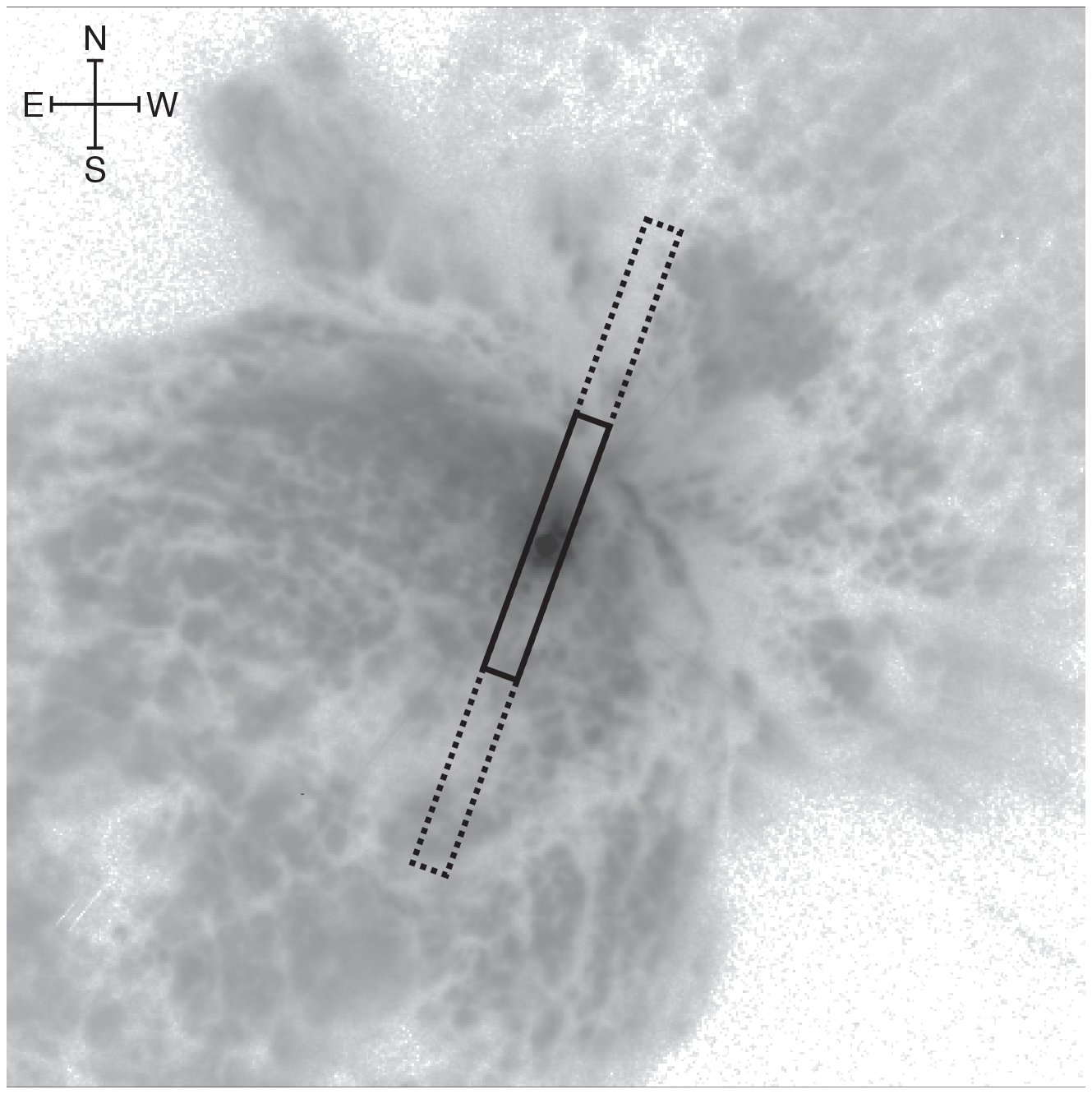}
  \includegraphics[angle=0, scale=0.315]{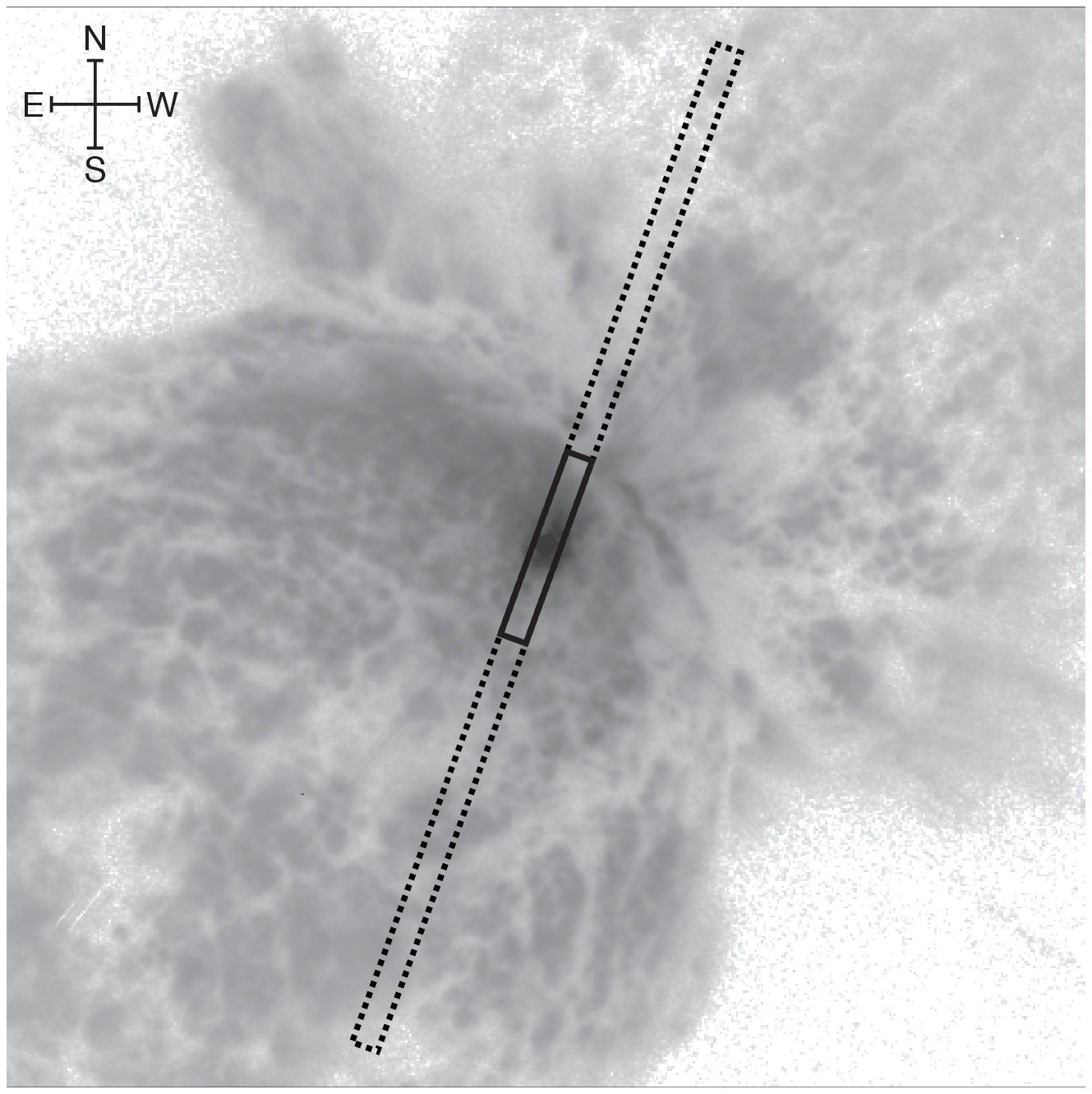}
  \caption{Image (12\arcsec $\times$ 12\arcsec) obtained with \hst\ Advanced Camera for Surveys  (GO program 9420; PI: K. Davidson). The UVES blue arm (0\farcs4 $\times$ 7\farcs6) and red arm (0\farcs3 $\times$ 11\farcs8) apertures are pictured (dashed) on the left and right images, respectively. In both figures the extracted parts of the apertures (solid) are depicted as used for the 2002 December 12 observation: 3\farcs00 in the blue arm and 2\farcs04 in the red arm. The cross-hair is 1\arcsec\ in both directions. \label{nielsen_f1} }
\end{figure}

\citet{Weis05a, Bomans05a} did the initial analysis of \ec's  \vlt/UVES spectra and noted ejecta absorption lines including the \ion{Na}{1}, \ion{K}{1} and \ion{Ca}{2} doublets. The \vlt/UVES data were obtained during the \textit{"\ec\ Campaign with UVES at the ESO VLT"} which is a long term project for spectroscopic monitoring of \ec\ with \vlt/UVES in two slit positions. The first slit position is centered on the star itself and provided the data used by \citet{Weis05a, Bomans05a} and is the base for the investigation presented in this paper. The second slit position is offset towards the south-east lobe of the Homunculus nebula. \citet{Stahl05} used the scattered light spectrum at the center of this slit position in their analysis of the \ion{He}{2} \lm4686 line profiles. \\
\indent \vlt/UVES provides the necessary spectral resolving power for spectral investigations of the ejecta at longer wavelengths. The extended wavelength coverage includes species with transitions outside the \hst/STIS echelle spectral range, such as \ion{Sr}{2}, \ion{Sc}{2}, NH and CH$^+$. However, the \vlt/UVES spectrum is not easily compared to the \hst/STIS spectra, because of the added complication of ground-based seeing coupled with nebular emission and spatially extended wind features from \ec~A and B that enter the larger aperture.\\
\indent This paper presents the UVES spectrum between 3060 and 10,430~\AA, with line identifications of the nebular emission features, the P-Cygni wind lines and the ejecta absorption. The paper provides the necessary tools for further investigations of \ec's spectrum, including an atlas with line identifications in addition to an analysis of the scattered contribution to the UVES continuum. Section~\ref{obs} presents the observational  data used in the analysis. Section~\ref{spectra} discusses the spectra of the Weigelt condensations, the circumstellar material and the stellar wind. In Section~\ref{colden} we compare the column density measurements in the \vel{513} absorption component to understand the scattered light properties that strongly influence the \vlt/UVES spectrum.  The wavelength overlap between \vlt/UVES and \hst/STIS provides the means to investigate the UVES spectrum with respect to the scattered radiation from the central source and is invaluable for further investigations of the UVES spectrum. We discuss the wavelength region between 3060 and 3159~\AA, and derive a correction factor for the added scattered stellar radiation in the UVES spectrum. The correction factor will be used in an abundance analysis of the ejecta, including all observed species in absorption, that currently is being prepared by Kober et\,al. The online edition of this paper includes the entire \vlt/UVES spectrum with over 1500 line identifications. All wavelengths throughout the paper are in vacuum. 

\begin{deluxetable}{lcc}
\tabletypesize{\scriptsize} 
\tablecaption{\hst/STIS vs. \vlt/UVES \label{nielsen_t1}} 
\tablewidth{0pt} \tablehead{
\colhead{} &
\colhead{\hst/STIS\tablenotemark{a}} &
\colhead{\vlt/UVES\tablenotemark{b}} 
} 
\startdata
$\lambda$-coverage (\AA) & 2380 $-$ 3159 & 3060 $-$ 10,430\tablenotemark{c} \\
                                           &                         & 3060 $-$ 4900 (blue)\tablenotemark{d} \\
                                           &                         & 4900 $-$ 10,430 (red)\tablenotemark{d} \\ 
Aperture (\arcsec) & 0.3 $\times$ 0.2 & 0.4 $\times$ 7.6 (blue) \\ 
         & & 0.3 $\times$ 11.8 (red) \\
Resolving Power: & & \\   
Spectral ($\lambda$/$\Delta \lambda$) & 110,000 & 80,000 (blue) \\ 
         &                         & 100,000 (red) \\
Angular (\arcsec) & 0.07 & \phn0.39$-$1.67\tablenotemark{e}  
\enddata
\tablenotetext{a}{\hst/STIS MAMA spectrum is from \hst\ GO program 9973 (PI: K. Davidson) and is available at  \url{http://archive.stsci.edu/prepds/etacar/} or  \url{http://etacar.umn.edu/}.}
\tablenotetext{b}{\vlt/UVES spectra are obtained during program IDs: 070.D-0607, 071.D-0168, 072.D-0524, 074.D-0141, 077.D-0618, 
380.D-0036; PI: K. Weis.}
\tablenotetext{c}{Complete wavelength range for these observations. Small gaps in wavelength coverage due to spacing between the CCDs.} 
\tablenotetext{d}{Used wavelength range.} 
\tablenotetext{e}{Seeing-limited.}
\end{deluxetable}

\begin{deluxetable}{llcc}
\tabletypesize{\scriptsize} 
\tablecaption{Phase coverage for the used observational data. \label{nielsen_t2}} 
\tablewidth{0pt} 
\tablehead{
\colhead{Instrument}  &
\colhead{Observation Date} &
\colhead{JD\tablenotemark{a}} &
\colhead{Phase, $\phi$\tablenotemark{b}}
} 
\startdata
\vlt/UVES\tablenotemark{c} & 1999 Dec 21  & 1534   & 0.363 \\   
&2002 Dec 7    & 2616   & 0.897 \\
&2002 Dec 12  & 2621   & 0.900 \\
&2003 Feb 14  & 2685   & 0.931 \\                           
&2003 May 29  & 2789   & 0.983\\
&2003 Jun 3    & 2794    & 0.985\\
&2003 Jul 5     & 2826   & 1.001 \\
&2004 Feb 20 & 3056    & 1.115 \\
&2005 Feb 12 & 3414 & 1.292 \\
&2005 Mar 19  & 3449   & 1.309 \\   
&2006 Apr  9    & 3835   & 1.500 \\
&2006 Jun 8     & 3895   & 1.529 \\ 
&&& \\   
\hst/STIS \tablenotemark{d} & 2004 Mar 6  & 3071 & 1.122  \\
 \enddata
\tablenotetext{a}{JD$-$2,450,000}
\tablenotetext{b} {Phase relative to X-ray minimum 1997.9604, JD2,450,799.792+2024$\times\phi$; \citet{Corcoran05a}.}
\tablenotetext{c} {The spectra used in this analysis is an extraction to include 85\% of the flux from the central source.}
\tablenotetext{d} {The spectrum used in this analysis is a 0\farcs0145 extraction centered on \ec.}
\end{deluxetable}

\section{Observations} \label{obs}
Coordinated, highly complementary, observations were made with the \hst/STIS and the \vlt/UVES before, during, and after the spectroscopic low-state centered on 2003.5. The STIS spectra were obtained at critical intervals and cover the UV spectral range with high spatial and spectral resolving power. The STIS echelle observation, used in this analysis, was obtained in 2004 March with the E230H grating setting, providing a 0\farcs070 spatial resolution with 0\farcs0145 spatial sampling and a spectral resolving power of 110,000 in the wavelength region 2380$-$3159~\AA. The area that is covered by the STIS slit changes as a function of the spacecraft roll-angle. For this particular observation, the \hst\ orientation angle permitted positioning of the 0\farcs3$\times$0\farcs2 aperture to include Weigelt~B and D in addition to \ec. The spectrum, used in this analysis, is an extraction of  seven 0\farcs0145  wide  high-resolution rows (7$\times$0\farcs0145=0\farcs102) from the \hst/STIS MAMA echelle image centered on \ec. The 0\farcs102 extraction minimizes contamination from the surrounding nebula and maximized the amount of information of the stellar spectrum. To avoid potential flux modulation, we used a spectral extraction that is slightly larger than what is needed based on the instrument's spatial resolving power. The \hst/STIS data was acquired from the MAST data archive and reduced with a non-standard IDL reduction software from the STIS instrumental development team \citep{Lindler99,Valenti02}. \\
\indent The UVES spectra cover 3060$-$10,430~\AA,  with small breaks due to physical gaps between the CCD detectors, with 80,000 and 100,000 spectral resolving power in the blue (\lm$_c$=3460~\AA) and red (\lm$_c$=8600~\AA), respectively.  Spectra were obtained frequently  across the 2003.5 low-state and with annual visits during the recovery period through 2006.4. All UVES high dispersion spectra were obtained with the same position angle (PA=160$^\circ$) as shown in Figure~\ref{nielsen_f1}. While seeing was recorded by the observers to range between  0\farcs39 to 1\farcs67, the recorded line by line spectra have significantly larger FWHM, likely due to the extended geometry of \ec.  The recorded \vlt/UVES spectra indicated significant larger stellar profiles along the slit compared to the observer-recoreded seeing measures. Therefore, we used a slit extraction with variable spatial width to all stellar flux down to 15 percent that of the peak flux. For the spectrum obtained in 2002 December 12 we used a 3\farcs00 and 2\farcs04 extraction of the blue and red aperture, respectively. For observations obtained at the other phases, extractions as large as $>$3\farcs75 were necessary to include the bulk of the flux from the star. The UVES data were reduced with pipeline methods. The resultant spatially resolved line-by-line spectra were produced by one of the authors (O. Stahl) with tailor made software to optimize the signal-to-noise in the overlap regions. The UVES spectra are not flux-calibrated. Comparisons between different spectra were done with the continuum normalized to unity. \\
\indent The UVES and STIS characteristics are summarized in Table~\ref{nielsen_t1} and the observational phase coverage of \ec's 5.54 year spectroscopic period is presented in Table~\ref{nielsen_t2}. Figure~\ref{nielsen_f1} shows \hst/Advanced Camera for Surveys optical images  (GO program 9420; PI: K. Davidson) with the UVES apertures overlaid, including the portions of the apertures for which spectra were extracted for the 2002 December 12 observation. \\
\indent  We compare the \vlt/UVES spectrum with the \hst/STIS E230H in the overlap region between 3060 and 3159 \AA\ to gain insight of what is included in the UVES aperture. We extend the analysis to the entire covered UVES spectrum. The complete spectrum is presented as an atlas with line identifications in the online edition of the paper.

\section{The \ec\ Spectrum} \label{spectra}
\begin{figure*}[ht]
\centering
\includegraphics[width=12cm, angle=0]{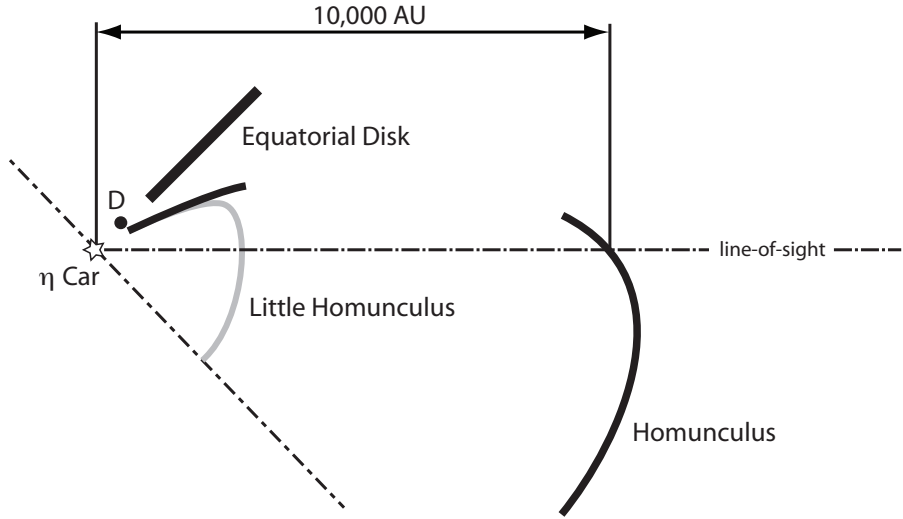}
  \caption{The current view of the \ec\ system. The UV/visible spectrum is dominated by spectral lines primarily originating from the dense wind of \ec~A. Narrow emission lines are observed from Weigelt condensations, represented in the figure by Weigelt~D. The line-of-sight from \ec\ intersects the ejecta, specifically  the Little Homunculus and the Homunculus. Consequently,  large parts of the observed spectrum are dominated by the ejecta absorptions. \label{nielsen_f2}}
\end{figure*}

\begin{figure*}[t]
\centering
\includegraphics[width=12cm, angle=90]{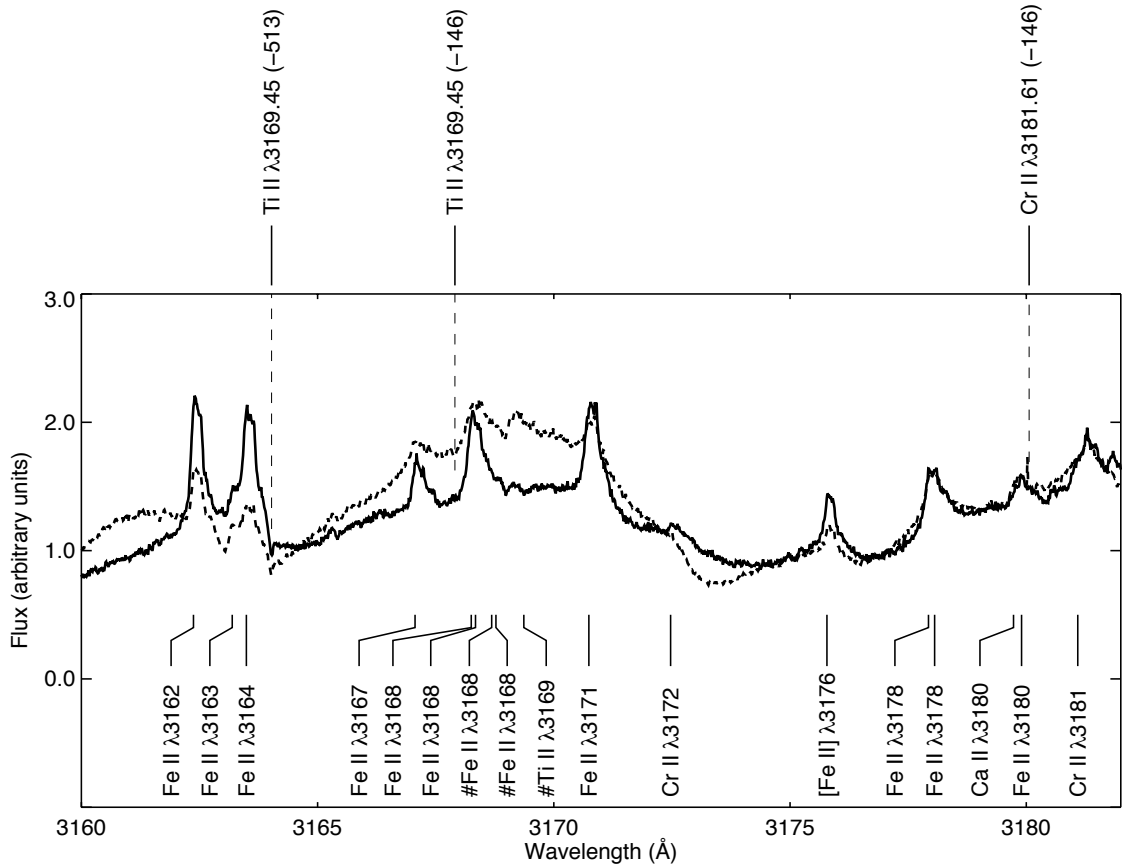}
  \caption{Extracted high-resolution spectrum obtained with \vlt/UVES (R$\sim$80,000) of \ec\ during the spectroscopic high-state (2002 December 12,  solid) and spectroscopic low-state (2003 July 5, dashed). Both spectra are dominated by the emissions from the Weigelt condensations and \ec~A's stellar wind lines. Narrow absorption features originating from the circumstellar material, are observed at $-$146 and \vel{513}. Identifications for the ejecta absorption lines are in the top of the figure with wavelength and radial velocity. Stellar wind lines and emission from the Weigelt condensations are identified in the bottom of the figure with wavelength. The wind lines are marked with \#.  All identified lines are presented in Table~\ref{nielsen_t3}. All wavelengths are in vacuum. \label{nielsen_f3}}
\end{figure*}

The \ec\  spectrum is modulated by the 5.54 year spectroscopic period.  During the greater part of the period the system is influenced by radiation from the hotter \ec~B and is referred to as the spectroscopic high-state. The spectroscopic high-state is characterized by a higher level of  ionization particularly in the Weigelt condensations and the circumstellar material. Across  \ec~B's periastron passage, the system is dominated by radiation from \ec~A and exhibits lower ionization. The periastron passage lasts for a few months and is often called the spectroscopic minimum or low-state. \\
\indent The spectra observed with \vlt/UVES are seeing-limited and include extended stellar wind lines plus narrow line emission from the surrounding nebula, from the Weigelt condensations in particular. At wavelengths longward of 6250~\AA, the Telluric O$_2$ and H$_2$O bands heavily contaminate the spectrum and significantly impair the analysis of the velocity variability and the changes in integrated flux of the stellar wind lines. In the following sections we discuss the origins of the identified lines in the spectrum, including the Weigelt condensations, the stellar wind and the circumstellar material. Figure~\ref{nielsen_f2} shows the current view of the system including how our sightline intersects the ejecta. Figure~\ref{nielsen_f3} shows an example of the UVES spectrum with identifications. It is  a sample of the full spectrum that appears in the online edition of the paper. More information about the identified transitions in Figure~\ref{nielsen_f3} is presented in Table~\ref{nielsen_t3}.  The complete figure set, available online,  is a comparison between a spectrum obtained during the spectroscopic high-state (2002 December 12, $\phi$=0.90) and one recorded during the low-state (2003 July 5). The two spectra are compared over the entire \vlt/UVES  wavelength region (3060 $-$ 10,430 \AA). A reference key for the online atlas is provided in the end of the paper (Figure~11)

\begin{deluxetable*}{llcccrcc}
\tabletypesize{\scriptsize} 
\tablecaption{Identifications in the spectral interval 3160$-$3182 \AA.\label{nielsen_t3}} 
\tablewidth{0pt} \tablehead{
\colhead{$\lambda_\mathrm{obs}$\tablenotemark{a}} &
\colhead{} &
\colhead{$\lambda_\mathrm{lab}$\tablenotemark{b}} &
\colhead{} & 
\colhead{Velocity\tablenotemark{c}} &
\colhead{E$_\mathrm{low}$\tablenotemark{d}} &
\colhead{} &
\colhead{} \\
\colhead{(\AA)} &
\colhead{Line ID} &
\colhead{(\AA)} &
\colhead{Comment} & 
\colhead{(\kms)} &
\colhead{(cm$^{-1}$)} &
\colhead{log\,$gf$} &
\colhead{Reference\tablenotemark{e}} 
} 
\startdata
  3162.37 &        \textup{Fe}\,\textsc{ii}      $ (d^7\,a^4P_{3/2} - 4p\,z^4F_{3/2})$  &   3162.89 &    em  &  \phn$-$49 &  13905 &  $-$2.84 &     R98a \\
   3163.19 &       \textup{Fe}\,\textsc{ii}       $(4s\,c^2G_{7/2} - 4p\,y^2G_{7/2})$  &   3163.71    &   em  &  \phn$-$49 &  33501 &  $-$1.18 &     R98a \\
   3163.49 &       \textup{Fe}\,\textsc{ii}       $(d^7\,a^4P_{5/2} - 4p\,z^4F_{5/2})$  &   3164.01    &   em  &  \phn$-$49 &  13474 &  $-$2.85 &     R98a \\
   3164.03 &       \textup{Ti}\,\textsc{ii}        $(d^3\,b^4F_{9/2} - 4p\, z^4D_{7/2})$  &   3169.45    &   cs  & $-$513 &   1216 &  $-$0.14 &     P01b \\
   3167.06 &       \textup{Fe}\,\textsc{ii}       $(4s\,a^4P_{5/2} - 4p\,z^4D_{3/2})$  &   3167.58    &   em  &  \phn$-$49 &  13474 &  $-$3.08 &     R98a \\
   3167.91 &       \textup{Ti}\,\textsc{ii}        $(d^3\,b^4F_{9/2} - 4p\,z^4D_{7/2})$  &   3169.45    &   cs  & $-$146 &   1216 & $-$0.14 &     P01b \\
   3168.25 &       \textup{Fe}\,\textsc{ii}       $(4s\,b^2G_{7/2} - 4p\,z^2G_{7/2})$  &   3168.77    &   em  &  \phn$-$49 &  30764 &  $-$0.72 &     R98a\\ 
   3168.34 &       \textup{Fe}\,\textsc{ii}       $(4s\,b^4D_{5/2} - 4p\,x^4D_{7/2})$  &   3168.86    &   em  &  \phn$-$49 &  31388 &  $-$2.66 &     R98a \\
   3168.69 &       \textup{Fe}\,\textsc{ii}       $(4s\,b^2G_{7/2} - 4p\,z^2G_{7/2})$  &   3168.77   & wind  &  \phn\phn$-$8 &  30764 &  $-$0.72 &     R98a \\
   3168.78 &       \textup{Fe}\,\textsc{ii}       $(4s\,b^4D_{5/2} - 4p\,x^4D_{7/2})$  &   3168.86    & wind  &  \phn\phn$-$8 &  31388 &  $-$2.66 &     R98a \\
   3169.37 &       \textup{Ti}\,\textsc{ii}        $(d^3\,b^4F_{9/2} - 4p\,z^4D_{7/2})$  &   3169.45    & wind  & \phn\phn$-$8 &   1216 &  $-$0.14 &     P01b \\
   3170.74 &       \textup{Fe}\,\textsc{ii}       $(d^7\,a^4P_{3/2} - 4p\,z^4D_{1/2})$  &   3171.26    &   em  & \phn$-$49 &  13676 &  $-$2.61 &     R98a \\
   3172.47 &       \textup{Cr}\,\textsc{ii}       $(4s\,a^2P_{3/2} - 4p\,z^2P_{1/2})$  &   3172.99    &   em  &  \phn$-$49 &  35356 &  $-$0.53 &      K88 \\
   3175.78 &       [\textup{Fe}\,\textsc{ii}]     $(4s\,a^6D_{9/2} - 4s\,b^4D_{7/2})$  &   3176.30    &   em  &  \phn$-$49 &      0 &  $-$8.57 &      G62 \\
   3177.93 &       \textup{Fe}\,\textsc{ii}       $(4s\,b^4D_{7/2} - 4p\,x^4D_{7/2})$  &   3178.45    &   em  &  \phn$-$49 &  31483 &  $-$0.90 &     R98a \\
   3178.06 &       \textup{Fe}\,\textsc{ii}       $(4s\,b^4D_{1/2} - 4p\,y^4D_{1/2})$  &   3178.58    &   em  &  \phn$-$49 &  31368 &  $-$2.92 &     R98a \\
   3179.73 &       \textup{Ca}\,\textsc{ii}       $ (4p\,^2P_{3/2} - 4d\,^2D_{5/2})$  &   3180.25    &   em  &  \phn$-$49 &  25414 &   \phs0.51 &      B72 \\
   3179.90 &       \textup{Fe}\,\textsc{ii}       $(4s\,c^2D_{5/2} - 4p\,y^2F_{7/2})$  &   3180.42   &   em  &  \phn$-$49 &  38164 &  $-$0.87 &     R98a\\ 
   3180.06 &       \textup{Cr}\,\textsc{ii}       $(d^5\,a^4G_{11/2} - 4p\,z^4F_{9/2})$  &   3181.61    &   cs  & $-$146 &  20512 &  $-$0.20 &      N06 \\
   3181.09 &       \textup{Cr}\,\textsc{ii}       $(d^5\,a^4G_{11/2} - 4p\,z^4F_{9/2})$  &   3181.61    &   em  &  \phn$-$49 &  20512 &  $-$0.20 &  N06   
\enddata
\tablecomments{CS, EM, WIND denotes the nature of the spectral line in the atlas, accordingly. CS is for circumstellar absorption lines, EM is emission from Weigelt~D and WIND is stellar wind emission. Non-referenced data are from R.~Kurucz's Atomic Line Database (\url{http://cfa-www.harvard.edu/amdata/ampdata/kurucz23/sekur.html}). A portion of the online table is shown here for guidance regarding its content.}
\tablenotetext{a}{Calculated observed vacuum wavelength based on the heliocentric velocity.}
\tablenotetext{b}{Vacuum rest wavelength.}
\tablenotetext{c}{Heliocentric velocity.}
\tablenotetext{d}{Energy for the transitions lower state.}
\tablenotetext{e}{Reference for log\,$gf$.}
\tablerefs{
\textbf{B72} $-$ \citet{Black72},  
\textbf{G62} $-$ \citet{Garstang62a},
\textbf{K88} $-$ \citet{Kurucz88},
\textbf{N06} $-$ \citet{Nilsson06a},
\textbf{P01b} $-$ \citet{Pickering01b},
\textbf{R98a} $-$ \citet{Raassen98a}
}
\end{deluxetable*}

\begin{figure*}
\centering
  \includegraphics[width=12cm]{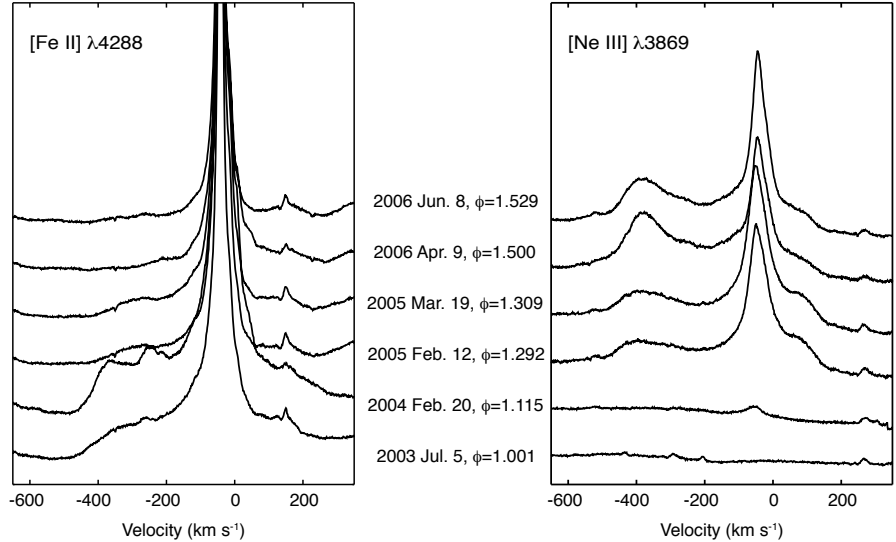}
  \caption{The recovery after the periastron passage of the emission lines from the Weigelt condensations  and the surrounding nebula. The transitions from higher ionization states, represented by [\altion{Ne}{iii}] \lm3869, are absent during the spectroscopic low-state ($\phi$=1.0) and reappear when the ionizing flux from \ec~B again penetrates \ec~A's wind. The spectral lines with lower ionization such as [\altion{Fe}{ii}] \lm4288, are less dependent on the external UV radiation and show less change in strengths over the 5.54 year spectroscopic period. Note the emission feature at \vel{400}. This component is strongest for the low ionization features at $\phi$=1.115 while the high ionization lines peak at apastron ($\phi$=1.500). The plots are offset vertically with increasing phase to show changes in the broad emission lines.  \label{nielsen_f4}}
\end{figure*}

\subsection{The Weigelt Condensations}
\citet{Zethson01b} analyzed the Weigelt~D emission spectrum  for the wavelength region covered by \hst/STIS CCD (1600$-$10,300 \AA, R$\sim$8,000). We used the work by \citeauthor{Zethson01b} as a guide to identify the narrow emission lines from the Weigelt condensations in the UVES spectrum. The  majority of the narrow lines observed with \hst/STIS are also observed in the UVES spectrum. However, weak lines are overpowered by the much strong stellar radiation in the \vlt/UVES spectrum. We note that the narrow component  of the emission lines recorded with high-resolution ($R\sim$80,000) \vlt/UVES are narrower than seen in the high-spatial, but lower spectral, resolution \hst/STIS spectra.   \\ 
\indent The most prominent nebular features from the Weigelt condensations during \ec's spectroscopic high-state are the \lya-pumped fluorescent lines in \ion{Fe}{2}, especially \ion{Fe}{2} $\lambda\lambda$2507, 2509 \citep{Johansson03}. Transitions from  higher ionization stages, such as Fe$^{2+}$, Fe$^{3+}$, Ne$^{2+}$, were also identified in the spectrum.  The radial velocities for the spectral lines in the spectrum of the Weigelt D condensation were determined by \citet{Nielsen07a} to be $-$47 \kms\ for the \lya-pumped lines and $-$49 \kms\ for the allowed transitions. These derived radial velocities are consistent with previous results from \citet{Zethson01b} and are used to identify the emission features in the \vlt/UVES spectrum. \\
\indent The \lya-pumped fluorescent lines and the forbidden lines of the highly ionized species are present during most of the 5.54 year cycle but disappear for several months during \ec's low-state, coinciding with \ec~B's periastron passage. The spectral variability is attributed to the UV radiation field from the hot \ec~B \citep{Verner05a}. During the periastron passage, the dense wind of \ec~A traps the FUV radiation from \ec~B in all directions. Figure~\ref{nielsen_f4} shows the differences in variability for the moderately ionized species, represented by [\ion{Fe}{2}] \lm4288 (\ion{Fe}{1} IP 7.9~eV), and the highly ionized [\ion{Ne}{3}] \lm3869 (\ion{Ne}{2} IP 40.9~eV) during the spectroscopic low-state and the recovery period. The highly ionized species are excited with radiation from \ec~B and, consequently, the most dramatic difference, as shown in the figure,  is the absence of the [\ion{Ne}{3}] \lm3869 at \vel{47} beginning at phase\footnote{phase relative to X-ray minimum 1997.9604, JD2,450,799.792+2024$\times\phi$; \citet{Corcoran05a}.}, $\phi$= 1, when the far-UV radiation from \ec~B is cut off.  The  [\ion{Fe}{2}] \lm4288 depends upon mid-UV radiation and can be excited in the dense wind of \ec~A. The narrow nebular emission shows little correlation with \ec's spectroscopic period. However, after the periastron passage ($\phi$=1.115), a broad blue shifted emission feature appears at \vel{400}. By contrast the  [\ion{Ne}{3}] \lm3869 is absent at $\phi$=1.0 with only a weak narrow component at $\phi$=1.115. As the blue shifted component for [\ion{Fe}{2}] \lm4288 fades, the [\ion{Ne}{3}] \lm3869 grows stronger through $\phi$=1.529. The blue shifted condensation that is responsible for the emission component, is again being ionized by the far-UV radiation from \ec~B. \\
\indent \citet{Zanella84a} noticed broad shoulders, blue-shifted up to 600 \kms, of emission lines in their observed spectrum between 1981 and 1983. They concluded the components originated from a dense shells excited by the UV radiation from the hot core. The \hst/STIS CCD spectrum, with 0\farcs1 resolution, do not always show these broad shoulders since the STIS aperture, with orientation, samples a only thin slice of the wind structure. By comparing \hst/STIS observations observed with different position angle and phase, we get indications of the origin of the broad emission components. A simple explanation for this emission is that it originates from the outer regions of the wind-wind interface. This would lead to a spatially resolved blue-shifted emission in highly excited lines, such as  [\ion{Ne}{3}] \lm3869, in a symmetric shell that is photo-ionized by \ec~B. In contrast, the less excited lines, for example  [\ion{Fe}{2}] \lm4288, are predominantly excited by the radiation from \ec~A.    

\subsection{The Ejecta Absorption}
The ejecta, the Homunculus and the Little Homunculus, are oriented such that their walls intersect our sightline towards \ec, leading to a major portion of the stellar radiation being absorbed (see Figure~\ref{nielsen_f2}). Consequently, \ec's spectrum in line-of-sight is defined by the broad wind lines from \ec~A with nebular emission and circumstellar absorption superimposed. \citet{Gull06a} identified over 30 velocity components in the near-UV spectrum where two components, at $-$146 and \vel{513}, are isolated and distinguishable from other velocity components. The Little Homunculus (\vel{146}) is observed in absorption from metastable energy levels in mainly singly-ionized iron-group elements \citep{Gull05a, Gull06a}. Their energy levels are populated by photo-excitation and thermal collisions. The Homunculus (\vel{513}) spectrum is abundant in narrow absorption lines in neutral and singly-ionized metals plus several diatomic molecules \citep{Verner05b}. The \vel{513} spectrum is strong in the \hmol\ Lyman bands located short-ward of 1650~\AA\ \citep{Nielsen05a}. These transitions are from states high up in the \hmol\ energy level system not normally populated in a cold gas. With a strong background UV source, the \hmol\ molecules can be excited followed by de-excitation to high vibrational and rotational states in the ground electronic state. The H$_2$ lines vary in absorption depth with spectroscopic phase, which is in agreement with changing UV radiation from the binary system. Other molecules, for example, \textup{CH} and \textup{OH} are present in the \hst/STIS spectrum, but no CO is observed. Additional molecules are observed in the \vlt/UVES spectrum at longer wavelengths, such as \textup{NH} and \textup{CH$^+$} \\
\indent  \citet{Gull05a} presented preliminary results for the abundance analysis of the \ec\ ejecta, based on  \ion{Fe}{2} and \ion{Ti}{2}  column densities in the $-$146 and \vel{513} velocity components, respectively. They derived column densities using a standard curve-of-growth method, where the results were used as input data for {\sc cloudy} modeling. {\sc cloudy} is a photo-ionization code where the energy level populations are calculated with a statistical equilibrium approach \citep{Ferland98a}. A  Kurucz model atmosphere \citep{Kurucz88} with $T_\mathrm{eff}$=15,000~K and $L$=10$^{40}$ ergs\,s$^{-1}$, with carbon and oxygen 100 times depleted and nitrogen and helium 10 times enhanced, was used to mimic  the radiative flux from \ec~A. All other metals were assumed to have solar abundances. The absorption at \vel{146} was determined to originate in a gas located 1300 AU from the radiative source, with a variable excitation temperature, $T_e$=4700$-$5750~K, dependent on the spectroscopic phase. The absorption at \vel{513} is located at $\sim$10,000~AU from the stellar system and is much cooler, $T_e$=760~K. However, modeling of transitions in \textup{CH} and \textup{OH} yield a much lower temperature \citep[30~K,][]{Verner05b}, indicating formation of atomic and molecular absorption in two spatially separated regions. No measurable difference in velocity between the molecular and atomic absorption lines is observed. The absorption at \vel{513} does not show any significant variability in atomic absorptions with phase. However, the strongly phase dependent \hmol\ fluorescent spectrum nearly disappears during the periastron passage when the UV radiation from the central source is cut off. A more comprehensive study covering all observed species in the \hst/STIS and the \vlt/UVES spectra is underway (Kober et\,al. in prep). The results from the modeling by \citet{Gull05a} are summarized in Table~\ref{t4}.\\
\indent The larger \vlt/UVES aperture includes contributions from additional nebular structures besides the radiation from the extended, dust scattering stellar wind lines and the narrow emission from the Weigelt condensations. The narrow line nebular absorptions as seen in the \hst/STIS UV spectrum extends into the \vlt/UVES spectral regions. Many absorption lines at \vel{513} and some at \vel{146} are still identified and measured.  \vlt/UVES provides high dispersion spectra to 10,430~\AA and thus the opportunity to investigate additional atomic and molecular species which are not accessible with the high spectral resolution \hst/STIS MAMA. The \hst/STIS CCD spectrum with low spectral resolution, extends beyond 10,000 \AA\ but does not show any signs of absorption features from the circumstellar shells. For the \vel{513} component associated with the Homunculus, UVES provides information about atomic absorption lines in \ion{Sr}{2}, \ion{Sc}{2}, \ion{K}{1}, \ion{Ca}{1}, \ion{Ca}{2}, \ion{Na}{1}, \ion{Cr}{1}, \ion{Ti}{2}, \ion{V}{2}, \ion{Ni}{1}, \ion{Fe}{1} and molecular lines in {\textup CH}, {\textup CH$^+$} and  {\textup NH}. The absorption spectrum is characterized by allowed transitions which are easier to interpret, regarding populations of the energy levels. Abundance analysis using absorption spectra provide more accurate results due to better known atomic/molecular data.

\begin{deluxetable}{lcc}
\tabletypesize{\scriptsize} 
\tablecaption{Ejecta parameter in part from \citet{Gull05a} \label{t4}} 
\tablewidth{0pt} \tablehead{
\colhead{  } &
\colhead{$-$513 \kms} &
\colhead{$-$146 \kms}
}\startdata
$T_{max} (K)$  &  760   &     6400\tablenotemark{b} \\
$T_{min}  (K)$  &  760\tablenotemark{c}   &     5000\tablenotemark{b} \\
Ionization & Neutral,  Singly &  Singly \\ 
Distance (AU)  &  10,000 & 1300 \\
$\rho_e$ (cm$^{-3}$) &  Ê10$^7$    &   10$^7$$-$10$^8$ \\
$b$ (\kms)                    &  2.1$-$2.5  &   5.5$-$8.0  \\
$v_{t}$\tablenotemark{a} (\kms) & Ê 1.4$-$1.7  &  \phn3.8$-$5.6
\enddata
\tablenotetext{a} {The $b$-value is the Doppler width of the spectral line. In this definition of $b$, $v_t$ is the one-dimensional rms value for a gaussian spectrum of internal trubulent velocities, the full three dimensional turbulent velocity is thus $\sqrt{3}$ $v_{t}$.}
\tablenotetext{b} {The temperatures have been revised with oscillator strengths from \citet{Raassen98b} yielding $T_{min}$=4700~K and  $T_{max}$=5750~K. }
\tablenotetext{c}{The temperature may change between the high and low states, but the difference is within the measurement error. The accuracy of the $T_\mathrm{min}$ is affected by blends due to additional velocity components that appear during the spectroscopic low-state. }

\end{deluxetable}

\subsection{Eta Car's Wind spectrum}
\begin{figure*}
\centering
  \includegraphics[width=12cm]{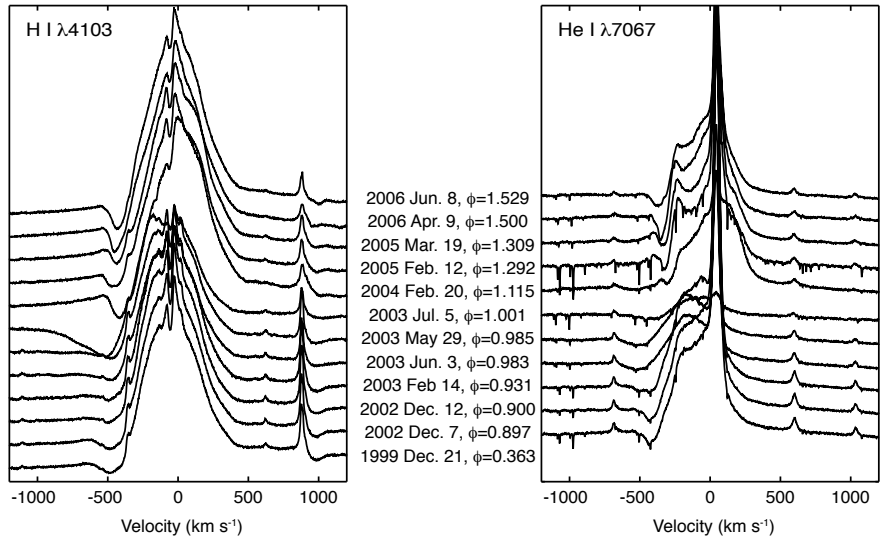}
  \caption{A comparison of the \altion{He}{i} \lm7067 originating in the wind-wind interface region with  \altion{H}{i} \lm4103 (H$\delta$) formed in \ec~A's wind. The Weigelt component dominates the emission of the \altion{He}{i} line, but with an underlying complex structure formed in the two arms of the bow-shock. The absorption component of the P-Cygni shaped line show a very abrupt shift at periastron. A similar, but less dramatic, shift is observed in H$\delta$, indicating the line is formed in a larger volume and further away from the radiative source.}
\label{nielsen_f5} 
\end{figure*}

The spectral crowding of absorption features complicates observations of the wind lines in the UV. The ejecta absorption, especially by the iron-group elements, blanket the spectrum and impairs the measurements of the wind lines and their time variability. At wavelengths longward of 3000 \AA\ the line density decreases and the wind lines, primarily from \ec~A, are observable. Most wind line profiles appear with a broad ($\pm$550 \kms) profile, often with P-Cygni absorption, plus a narrow ($\pm$15 \kms\ centered at \vel{47}) emission component, originating from the Weigelt condensations. The presence of the narrow emission from the Weigelt condensations  impairs the analysis of the velocity variation and the variability of the emission and absorption components of the wind lines. Many wind lines show significant variations in velocity and integrated flux over the 5.54 year spectroscopic period. The line variability is dependent on where the line is formed in the stellar wind. By investigating the phase dependence, information regarding the wind ionization structure and the influence of the companion star can be extracted. The wind spectrum between 3060 and 10,430~\AA\ is dominated by the hydrogen Balmer and Paschen series. The hydrogen lines are predominantly formed in \ec~A's wind with terminal velocity of $\sim$550 \kms. Their absorption profiles show small velocity variations, which likely are tied to the orbital motion of \ec~A. The weaker \ion{He}{1} line profiles show more pronounced velocity variations as observed by \citet{Nielsen07b}. The \ion{He}{1} lines originate from states high up in the energy level system ($<$150,000 \cm) and are normally not present in a 15,000~K stellar spectrum. In the \hst/STIS spectrum the \ion{He}{1} lines show complex line profiles with multiple emission and absorption components. The lines were, based on their presence,  velocity variation and line profile, determined to be formed in the wind-wind interface region between the two massive objects. With a lower spatial resolving power the multi-component structure is damped in the UVES spectrum leading to a smooth line profile.  \citet{Nielsen07b} used the \hst/STIS observed P-Cygni absorption to derive a radial velocity curve. However, the relation between the derived velocity and the motion of the companion star is difficult to understand. The changes in the UVES observed line profiles and their variability over the spectroscopic period are shown in Figure~\ref{nielsen_f5} for the \ion{H}{1} \lm4103 and \ion{He}{1} \lm7065. Other species with strong prominent wind spectrum are Si$^+$ and Fe$^+$, which due to their ionization potentials, have lines primarily formed \ec~A's extended stellar wind.

\section{Column Density Measurements}\label{colden}
\begin{figure}[ht]
\centering
  \includegraphics[width=8cm]{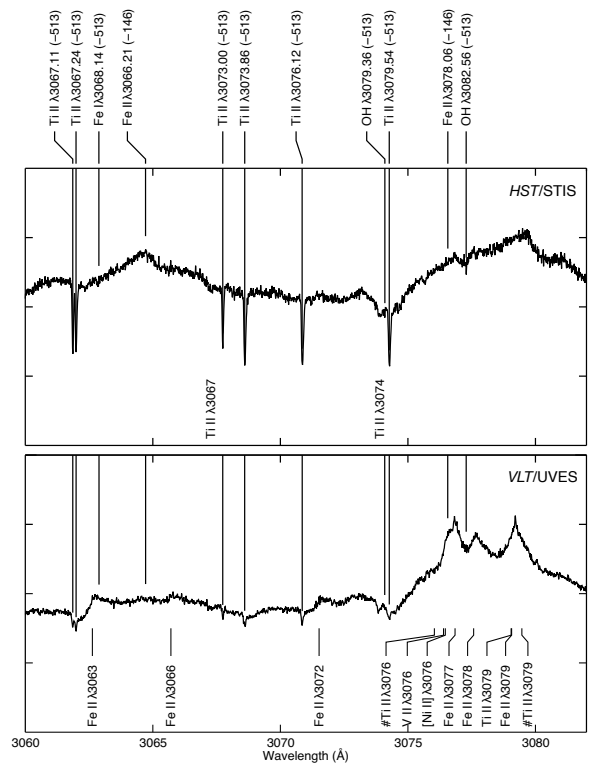}
  \caption{Comparison of the \hst/STIS (top) and \vlt/UVES  (bottom) spectra in the overlap region. The ejecta absorption spectrum in this wavelength range is dominated by transitions in \altion{V}{ii} and \altion{Ti}{ii}. The emission from the Weigelt condensations does not influence the \hst/STIS spectrum but is included in the \vlt/UVES spectrum. The \vlt/UVES spectrum is recorded in 2004 February 20 and the \hst/STIS spectrum is from 2004 March. All wavelengths are in vacuum.  \label{nielsen_f6}}
\end{figure}

The abundance analysis is dependent on  accurately measured elemental column densities. While the majority of the ejecta absorption lines useable for equivalent width measurements are in the \hst/STIS spectrum (2380$-$3159 \AA), some species are uniquely observed in the \vlt/UVES  spectral region (3159$-$10,430 \AA). Meaningful comparisons of column densities require correcting the measured equivalent widths for all observational effects. The \hst/STIS spectrum was obtained with near diffraction spatial resolution excluding emission from the surrounding nebula and the extended stellar wind. The seeing-limited spatial  resolution of \vlt/UVES required a larger aperture that includes the emission from the Weigelt condensations and most of \ec~A's extended wind. Hence, the stellar scattered by the ejecta and the extended wind contribute to the overall stellar continuum recorded by UVES. To compare the observed spectra and equivalent widths for lines that are recorded with \hst/STIS and \vlt/UVES, respectively, we used the small overlap in wavelength coverage between the two instruments (3060 to 3159~\AA).  \\
\indent The wavelength region 3060$-$3159 \AA\ has many \ion{Ti}{2} and \ion{V}{2} absorption lines. A subset of the overlap wavelength region is presented in Figure~\ref{nielsen_f6}. \ion{Ti}{2} was used by \citet{Gull05a} to derive the physical parameters for the Homunculus. The column density is derived from a standard curve-of-growth, i.e. the equivalent width as function of the transition strength. To accurately calculate the column density for an energy state, the transitions from this energy level must be unblended, have accurately determined atomic data, and be located in a wavelength region where the continuum can be properly set. At 3000~\AA\ the absorption line density is significantly lower, compared to the mid and far UV. The probability for blends decreases at longer wavelengths and makes the continuum placement easier, especially in the \hst/STIS high-resolution spectrum that is unaffected by the strong emission from the surrounding nebula. The continuum placement for the UVES spectrum is affected by the scattered emission features from the nebula and, at longer wavelengths, the atmospheric absorption bands. In the wavelength region between 3060 and 3159~\AA, the contribution from the narrow line atmospheric absorption is negligible and only a few lines are affected by the narrow emission from the Weigelt condensations.  Our analysis does not include absorption lines that are obviously blended or where the continuum level could not be set with a satisfactory accuracy. Our confidence in setting the continuum is comparable for both the STIS and the UVES spectra in the wavelength region, 3060$-$3159~\AA, and is helped by the narrow width of the ejecta absorption lines ($b$$\sim$2.5 \kms). The overlap region allows us to make a direct comparison of the equivalent widths for a few \ion{Ti}{2} lines measured in the UVES and STIS spectra. However, while using spectral lines in the overlap region to calculate a correction factor, additional \ion{Ti}{2} lines at wavelengths outside the overlap region are included  to derive and compare column densities. The \ion{Ti}{2} transition probabilities are, with a few exceptions, derived from measured branching fractions and lifetimes \citep{Pickering01b}, complemented with data from \citet{Fuhr88a}. 

\begin{figure}[ht]
\centering
  \includegraphics[width=8cm]{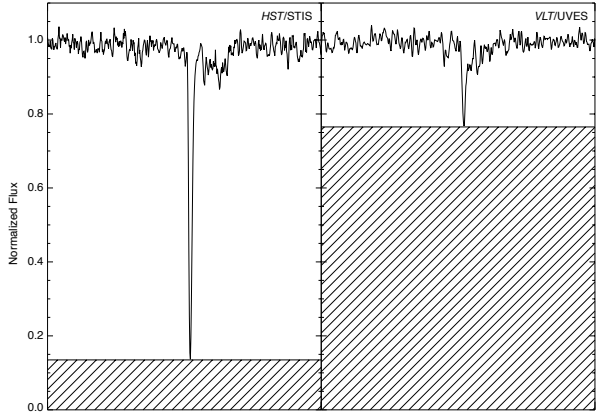}
  \caption{Comparison between the contribution from scattered radiation in the \hst/STIS (left: spectra obtained in 2004 March) and the \vlt/UVES  (right: 2004 February 20) spectra. The wavelength region is 3080$-$3087 \AA, including the ejecta absorption lines \altion{Ti}{ii} \lm3089 shifted to \vel{513}. The hashed area correspond to the contribution from scattered radiation.  \label{nielsen_f7}}
\end{figure}

\begin{figure}[ht]
\centering
    \includegraphics[width=8cm]{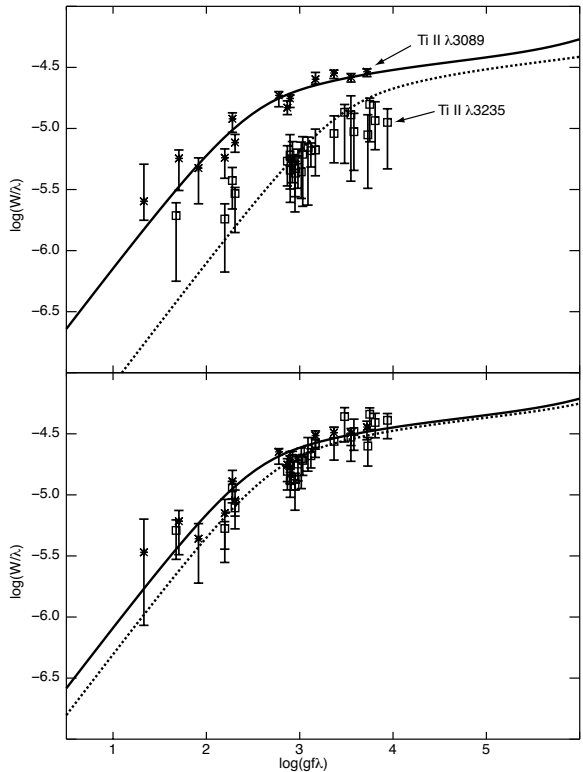}
   \caption{Comparison between curves-of-growth from data obtained from equivalent width measurements in the \hst/STIS (solid curve) and \vlt/UVES spectra (dashed curve), respectively. The \hst/STIS spectrum is from 2004 March 6, while the \vlt/UVES spectrum is from 2004 February 20. The upper panel show the curves-of-growth for the ground term $4s\,a\,^4\!F$ in \altion{Ti}{ii}. The column density from the STIS spectrum is nine times larger than corresponding values using the UVES spectrum. The lower panel show the curves-of-growth corrected for scattered radiation. The correction factor in the STIS spectrum is derived from the \altion{Ti}{ii} \lm3089, while \altion{Ti}{ii} \lm3235 was used for the UVES spectrum. The error bars represent the quality of the line fitting and the accuracy of the continuum placement. \label{nielsen_f8}}
\end{figure}

\begin{figure}[ht]
\centering
  \includegraphics[width=8cm]{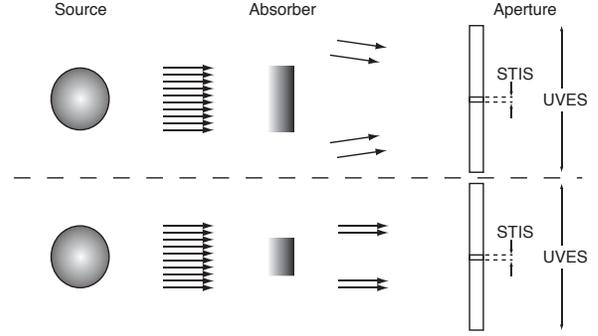}
  \caption{Illustration to the alternatives to explain the additional flux that enters the UVES aperture. The comparison of the UVES and STIS aperture sizes is based on the extracted parts for the spectra. In the blue spectral region the extracted part of the UVES aperture is 12$\times$0\farcs25 =  3\farcs00 and 7$\times$0\farcs0145 = 0\farcs1015 for STIS. Top: The absorber, the Homunculus covers the radiation source and the additional flux contribution that enters the UVES aperture comes from scattered star  light in the Homunculus itself.  Bottom: The Homunculus does not completely cover the source, and the full continuum is not absorbed. For the latter to be feasible the clumpiness of the Homunculus must be significant.   \label{nielsen_f9} }
\end{figure}

The absorption features recorded by \hst/STIS are systematically deeper than identical features recorded with \vlt/UVES. Hence, column densities derived from STIS spectra are larger than those from the UVES spectra. The \ion{Ti}{2} $\lambda$3089, shown in Figure~\ref{nielsen_f7}, is nearly saturated and falls close to the flat part of the curve-of-growth (see Figure~\ref{nielsen_f8}). However, it does not reach zero flux. The line profiles in the UVES spectrum do not show the same absorption depth, indicating a difference in detected continuum for UVES and STIS. The altered flux level may be caused by additional radiation entering the line-of-sight or instrumental effects. The latter is a combination of inter-order radiation and instrumental resolving power. The inter-order characteristics of the echelle spectrograph with a cross-dispersion grating can be examined using spectral cross-cuts of the detector image. For both instruments the inter-order contribution is small and corrected during the raw-data reduction process. \\
\indent  The modification of line profiles by limited spectral resolving power's was investigated through convolving a saturated test spectral feature with a point-spread-function of desired line width to observe how the spectral line depth changes. We estimated that \hst/STIS, with a spectral resolving power of R$\sim$100,000, changes the depth of the spectral feature by 5-10\%. This is in agreement with observations of the almost saturated \ion{Ti}{2} \lm3089 in the \hst/STIS spectrum. We conclude that the instrumental effects for STIS and UVES spectra are comparable in nature and do not explain the observed difference in absorption line depth.\\
\indent Saturated absorption lines, not reaching zero flux level, are observed in, for example, Seyfert galaxy spectra  \citep{Kraemer02}, where the absorber does not cover the entire radiative source. A similar effect is seen even if the absorbing cloud does cover the entire radiative source, but there is an added contribution to the continuum through reflected or scattered radiation (see Figure~\ref{nielsen_f9}). With a contribution to the continuum between the absorber and the observer, the additional radiation will cancel parts of the observed absorption and a saturated line would not reach zero flux at any wavelength. \\
\indent In the case of \ec\ and its ejecta, the extended wind and the ejecta contribute scattered radiation to the stellar continuum in the \vlt/UVES. The \vel{513} absorption is, however, originates in the Homunculus  located far from the radiative source. If the scattered radiation from the extended wind and the ejecta enters the light path before being absorbed in the Homunculus, the additional radiation will not alter the spectral line depth relative to the continuum.  

\begin{figure}[ht]
\centering
    \includegraphics[width=8cm]{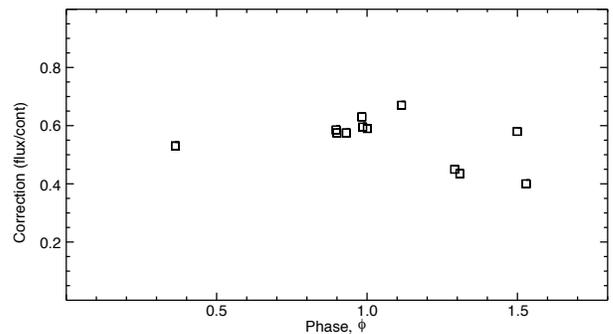}
   \caption{Phase dependance of the scattered radiation contributing to the continuum. The correction factor is calculated with \altion{Ti}{ii} \lm3235. The star was not centered in the aperture during the 2006 April 9 observation ($\phi$=1.500). Consequently, the measured contribution from scatted radiation is uncertain for this phase. The phase is relative to X-ray minimum 1997.9604, JD2,450,799.792+2024$\times\phi$; \citet{Corcoran05a} \label{nielsen_f10}}
\end{figure}

\indent The correction for the altered continuum is performed by estimating the offset using saturated lines in the spectrum. If the radiative source is entirely covered by the absorbing gas, then a saturated line is expected to reach zero flux at its line center. A few \ion{Ti}{2} lines in the spectral overlap region are close to the flat part of the curve-of-growth and, consequently, are assumed to be saturated. The \ion{Ti}{2} $\lambda$3089 is the best example of a saturated line in the overlap spectrum and is used to estimate the contribution of scattered radiation in the \hst/STIS spectrum. However, a few other lines appear to be better candidates in the UVES spectrum. We have chosen \ion{Ti}{2} \lm3235 to correct for scattered radiation in the UVES spectra. \ion{Ti}{2} \lm3235 is clearly on the flat part of the curve-of-growth (see Figure~\ref{nielsen_f8}) and has a well defined continuum. This spectral line is slightly blended by additional velocity component from the same transition. The contribution from the blending components to the line depth is small and the contribution to the correction factor is assumed to be negligible. Only a few lines are close to being saturated in the observed spectral region. Rather than calculate an average contribution factor from all nearly saturated lines, which would add a systematic error to the analysis, we have chosen to use one single line to calculate the correction. By adjusting the continuum in the UVES and STIS spectra, it is possible to compare column densities measured in the different wavelength regions observed by different instruments. We calculate a correction factor for the scattered radiation by subtracting the flux between zero flux and the measured line depth for a saturated line, according to Equation~\ref{nielsen_eq1}. 

\begin{equation}
\label{nielsen_eq1}
F_{c}=\frac{F}{F_{0}} - C_{f}. Ê
\end{equation}

Where $F$ is the pre-normalization flux level, $F_0$ is the defined  continuum level and $C_f$ is the correction factor.  If the correction factor is applied to the column density measurement made in the UVES and STIS spectra, respectively, then the results are in good agreement. Figure~\ref{nielsen_f8} shows how we have applied the correction factor for spectral lines in the STIS and UVES region for the $4s\,a\,^4F$ term in  \ion{Ti}{2}. The correction proved very effective in bringing the measured equivalent widths and derived column densities into agreement within the measurements errors. Note, to accurately calculate the influence from scattered nebular light, saturated spectral lines must be used. Otherwise, it is difficult to distinguish between contribution from scattered light and lack of absorbing material in the gas. An analysis using \ion{V}{2} at \vel{513} produced similar results. \\
\indent Correction of nebular, wavelength dependent, scattering properties would require a distribution of saturated spectral features throughout the wavelengths region. We do have that luxury, but must depend on lines in the blue portion of the spectrum. However, most absorption features are located in the near-UV wavelength region. For the absorption features at longer wavelength, the use of saturated \ion{Ti}{2} lines in the near UV only provides an estimate of the amount of scattered radiation entering the \vlt/UVES aperture at longer wavelengths.   \\
\indent The difference in scattered radiation observed in the STIS and UVES spectra, i.e. the contribution to the UVES continuum, varies with time. We find no correlation between added radiation with seeing conditions or zenith angle, but  the continuum offset may change with \ec's spectroscopic 5.54 year period, as shown in Figure~\ref{nielsen_f10}.  During the periastron passage the contribution from the scattered radiation appears to be greater compared with the rest of the spectroscopic period.  This phase coincides with the phase ($\phi$=1.0) when the radiation from the hot \ec~B  is increasingly obscured and the wind opacity is higher, indicating a more efficient scatterer. While the Homunculus is time independent in atomic absorption, exciting radiation from the central source is dependent on the orbital position of \ec~B.  Consequently, the variability in the continuum offset in the  UVES spectrum is more likely associated with the scattered radiation from the extended primary wind than photons scattered by the Homunculus.   If the radiation that modifies the continuum originates from the inner part of the nebula, then the absorber needs to be smaller than the radiative source. This is not fully in agreement with a Homunculus being a continuous absorber. For example,  \citet{Smith06a} depicted the  emission from the Homunculus as a thin smooth skin in H$_2$.  A clumpy Homunculus is in many ways in better agreement with previous studies. The presence of molecules and singly ionized metals in the same gas observed at the same velocity, can be explained with temperature and density stratification along the Homunculus surface.

\begin{figure*}
\centering
\figurenum{11}
  \includegraphics[width=12cm, angle=90]{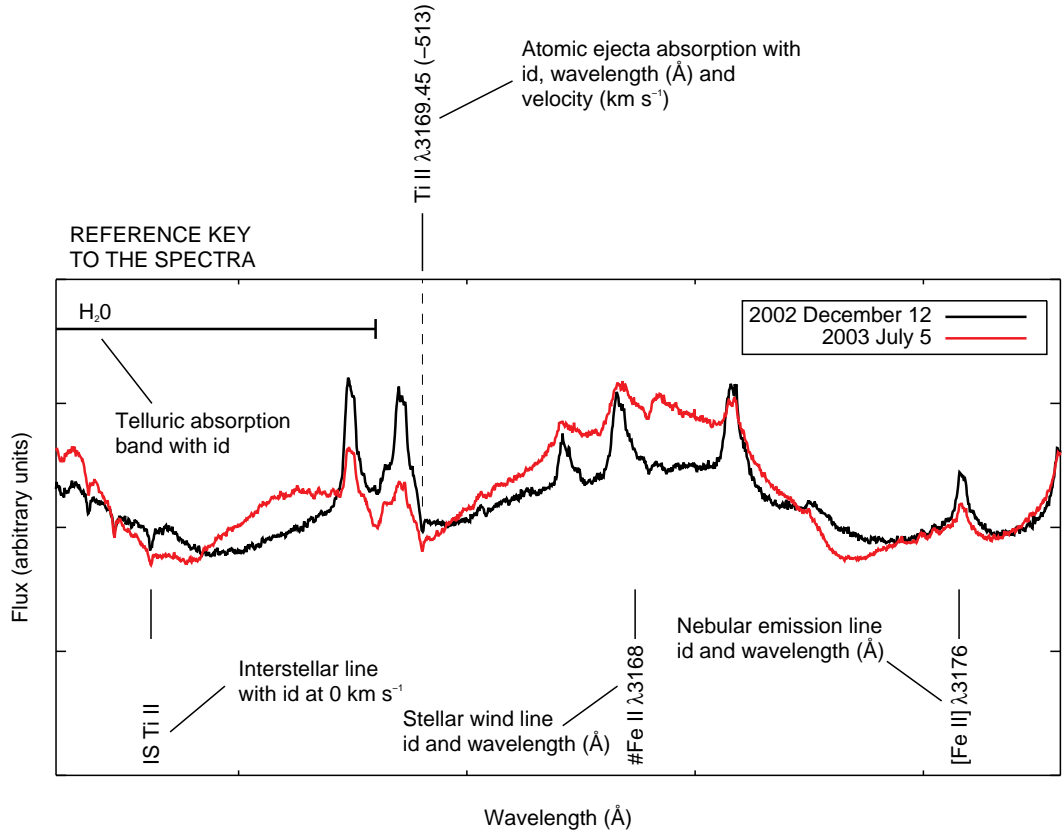}
  \caption{Reference key to all online only spectra. \label{nielsen_f11} }
  {\sc fig. set 11.} $-$ The \vlt/UVES spectra. [{\it See the electronic edition of the Supplement for Figs. 11.1$-$11.347}] 
\end{figure*}
Ê
\section{Summary}
\indent We have investigated the \vlt/UVES  spectrum between 3060 and 10,430~\AA\ and identified most spectral features. The spectrum is a composite of spectral lines from \ec~A's wind, emission from the surrounding nebula and the Weigelt condensations in particular, and absorption in the Homunculus and Little Homunculus. The \vlt/UVES spectra, from epochs $\phi$=0.900 and 1.001, are presented with over 1500 line identifications in the online edition of this paper.   \\
\indent The \hst/STIS and \vlt/UVES instruments have differences in spatial resolution and aperture size. We face the challenge to correct for the contribution of the nebular-scattered light entering the aperture. The scattered light will offset the continuum, which is a problem more pronounced for \vlt/UVES than for \hst/STIS spectra due to seeing and the larger aperture size. \\ 
\indent The \hst/STIS spectrum is obtained with the 0\farcs3 $\times$ 0\farcs2 aperture, but to minimize the contribution from the surrounding nebula a 7 pixel extraction (0\farcs102) was used in the analysis, providing critical separation of the stellar emission from dust scattered stellar radiation and nebular emission originating within the Homunculus. The UVES aperture is 0\farcs4 $\times$ 7\farcs6 for the blue spectrum, but we have used a subsection of the echelle image including flux from the central source down to a 15\% level. The limited spatial resolution in the, seeing dependent, ground-based UV spectra explains why the curves-of-growth from these two data sets appear to show different results. The  chaotic natures of the expanding ejecta within the Homunculus, the extended wind of the central source, and the many emission nebulosities lead to a velocity-smeared spectrum. \\
\indent The additional flux in the \hst/STIS can be explained either by the absorbing material in the Homunculus (at \vel{513}) being smaller than the radiative source in the center of the nebula, or by additional flux enters the aperture between the Homunculus and the observer. 
At the inferred distance of the Homunculus, $\sim$10,000 AU, the clumpy structure of the ejecta does not completely cover \ec's extended wind, based upon the differences in ejecta absorption profiles, yet the clumps in the Homunculus must be significantly larger than the 0\farcs1 that is covered by the \hst/STIS aperture, but smaller than the portion covered by \vlt/UVES. The structure of the Homunculus has significance for its gas-to-dust ratio and implications for its mass. The current gas-to-dust ratio in the Homunculus is very uncertain and an improved value would increase the derived value of the Homunculus mass.\\
\indent The amount of nebular/scattered light contribution, caused by scattered radiation, may change with orbital phase. This would make the correction even more problematic.  Phase dependence of the scattered radiation would suggest changes with in dust properties in the central region, likely due to new grain formation/destruction or modification by the UV radiation. A possible drop in the scattered light occurs about 1.7 years ($\phi$=0.3) after the spectroscopic low-states both in late 1999 and in 2005, however, the observed changes are on the same scale as the measurement errors. \\
\indent We have corrected for the additional flux in the UVES spectrum with saturated \ion{Ti}{2} lines in the blue and have shown that the correction puts the measurements from the \hst/STIS  and the \vlt/UVES spectra in agreement within the measurement errors. The correction of scattered radiation in UVES spectrum will be used to describe the  abundance characteristics of \ec's ejecta. That analysis is currently underway.

\acknowledgements
This analysis is based on observations collected at the European Southern Observatory, Paranal, Chile during programs: 070.D-0607, 071.D-0168,072.D-0524, 074.D-0141, 077.D-0618, 380.D-0036 and \hst\ GO program 9973.

\bibliographystyle{apj}
\bibliography{arXiv_ms74794}

\end{document}